# Two-Way Writing on Dirty Paper

Reza Khosravi-Farsani
Radio Communications Group
Iran Telecommunication Research Center (ITRC), Tehran, Iran
reza.khosravi@itrc.ac.ir

*Abstract*—In this paper, the Two-Way Channel (TWC) with Cannel State Information (CSI) is investigated. First, an achievable rate region is derived for the discrete memoryless channel. Then by extending the result to the Gaussian TWC with additive *interference noise*, it is shown that the capacity region of the later channel is the same as the capacity when there is no interference, i.e. a two-way version of Costa's *writing on dirty paper* problem is established.

*Keywords-Two-way channel; dirty paper channel ; channels with CSI.*

## I. INTRODUCTION

The Two-Way Channel (TWC) was introduced by Shannon [1], where an inner bound and also an outer bound on the capacity region was established. In 1984, Han [2] presented a general coding scheme for the discrete memoryless TWC, and derived an achievable rate region which exceeds the inner bound of Shannon in the general case. He also established the capacity region of the Gaussian TWC [2, Sec. VIII], and showed that for this channel feedback from outputs to inputs is completely ineffective. In fact, Shannon's inner and outer bounds for the Gaussian TWC are coincided, which yields the capacity. New outer bounds to the capacity region of TWC were obtained in [3].

Channels with Channel State Information (CSI) were also introduced by Shannon [4], where he established the capacity of a discrete single user channel with CSI causally known to the transmitter. For the case in which the transmitter has access to the CSI noncausally, the capacity was established in [5]. In 1983, in a famous paper named "writing on dirty paper", Costa [6] extended the result of [5] to the Gaussian channel with *additive interference* modeled as channel state, and showed that the capacity of the latter channel in which the interference is known noncausally to the transmitter is the same as the capacity when there is no interference, i.e. the known interference has no penalty on the capacity. The multiuser channels with causal and noncausal CSI have been also studied in several works (see [7] and the literature therein), specifically in [8] Costa's result for the dirty paper channel was generalized to some multiuser setups.

In this paper, we consider a TWC with CSI in which channel state is modeled by a 2-tuple $(S_1, S_2)$. Each user has access to the CSI *partially*, such that $S_i$ is known noncausally to the $i^{th}$ user, $i = 1,2$. We first derive an achievable rate region for the discrete memoryless channel. Then by extending the result to the Gaussian TWC with additive interference noise, we show that the capacity region of the later channel is the same as the capacity when there is no interference, i.e. a two-way version of Costa's writing on dirty paper problem is established.

## II. DEFINITIONS

In this paper, we use $X$, $x$ and $\mathcal{X}$ to denote a random variable (r.v.), its realization and range, respectively. The probability distribution function (p.d.f.) of a r.v. $X$ is denoted by $P_X(.)$, and $P_{X|Y}(.)$ stands for the conditional p.d.f. of $X$ given $Y$. The set of all jointly $\epsilon$-letter typical $n$-sequences $(x^n, y^n)$ with respect to the p.d.f. $P_{XY}(.)$ as defined in [9, Sec. 1.5], is denoted by $T_\epsilon^n(P_{XY})$. For a given sequence $x^n$, we use $T_\epsilon^n(P_{XY}|x^n)$ to denote the set of all $n$-sequences $y^n$ such that $(x^n, y^n) \in T_\epsilon^n(P_{XY})$.

*Definition:* A discrete memoryless TWC with CSI denoted by $(\mathcal{X}_1, \mathcal{X}_2, \mathcal{Y}_1, \mathcal{Y}_2, \mathcal{S}_1, \mathcal{S}_2, P_{S_1 S_2}, P_{Y_1 Y_2|X_1 X_2 S_1 S_2})$, is a channel with input alphabets $\mathcal{X}_1$ and $\mathcal{X}_2$ and output alphabets $\mathcal{Y}_1$ and $\mathcal{Y}_2$. Channel state is the 2-tuple $(S_1, S_2)$ which ranges over the set $\mathcal{S}_1 \times \mathcal{S}_2$ according to the $P_{S_1 S_2}$. Channel is described by the transition probability function $P_{Y_1 Y_2|X_1 X_2 S_1 S_2}(.)$. The system has been depicted in Fig.1. The state process of the channel is assumed to be independently identically distributed (i.i.d.), i.e. $Pr(s_1^n, s_2^n) = \prod_{t=1}^n P_{S_1 S_2}(s_{1,t}, s_{2,t})$, for $n \geq 1$, where $Pr(A)$ denotes the probability of occurrence of the event $A$.

*Encoding and decoding:* For the TWC with CSI $(S_1, S_2)$, in which $S_i$ is known noncausally to the $i^{th}$ user, $i = 1,2$, as depicted in Fig.1, a length-$n$ block code $\mathbb{C}^n(R_1, R_2)$ with two independent messages $W_i, i = 1,2$, where $W_i$ is uniformly distributed over the set $\mathcal{W}_i = \{1, \ldots, 2^{nR_i}\}$, consist of two sets of encoder functions $\{\varphi_{i,t}\}_{t=1}^n, i = 1,2$, defined as:

$$\varphi_{i,t} : \mathcal{W}_i \times \mathcal{S}_i^n \times \mathcal{Y}_i^{t-1} \to \mathcal{X}_i, \quad X_{i,t} = \varphi_{i,t}(W_1, S_i^n, Y_i^{t-1})$$

and two decoder functions $\xi_1(.)$ and $\xi_2(.)$ as:

$$\xi_1 : \mathcal{W}_1 \times \mathcal{S}_1^n \times \mathcal{Y}_1^n \to \mathcal{W}_2, \qquad \widehat{W}_2 = \xi_1(W_1, S_1^n, Y_1^n)$$
$$\xi_2 : \mathcal{W}_2 \times \mathcal{S}_2^n \times \mathcal{Y}_2^n \to \mathcal{W}_1, \qquad \widehat{W}_1 = \xi_2(W_2, S_2^n, Y_2^n)$$

The rate of the code is the pair $(R_1, R_2)$. A nonnegative pair $(R_1, R_2)$ is said to be achievable for the TWC with CSI, if for all sufficiently large $n$ there exists a length-$n$ block code $\mathbb{C}^n$ with the rate of $(R_1, R_2)$, for which the average error probability of decoding is arbitrary small. The capacity region of the channel is defined as the set of all achievable rates.

Note that in the above definition, one may restrict the encoders to depend only on the message and partial CSI intended to them. For this system called *restricted* TWC with CSI, the encoders functions $\{\varphi_{i,t}\}_{t=1}^n, i = 1,2$, are given by:

$$\varphi_{i,t} : \mathcal{W}_i \times \mathcal{S}_i^n \to \mathcal{X}_i, \quad X_{i,t} = \varphi_{i,t}(W_1, S_i^n)$$

*The Gaussian TWC with additive interference:* The Gaussian version of the channel depicted in Fig.1 is defined as the following:

$$Y_1 = aX_1 + bX_2 + S_2 + Z_1$$
$$Y_2 = cX_1 + dX_2 + S_1 + Z_2$$

(1)

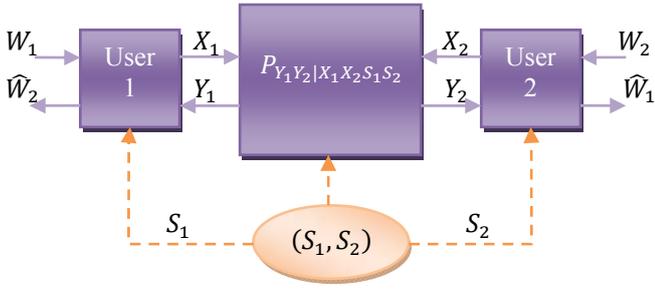

Figure 1. The TWC with CSI.

where $X_1, X_2$ and $Y_1, Y_2$ are real-valued input and output signals, respectively; $S_1, S_2$ are additive Gaussian interferences (potentially correlated) with zero mean and powers $P_{S_1}, P_{S_2}$, respectively, and $Z_1, Z_2$ are Gaussian noises (potentially correlated) with zero mean and powers $P_{Z_1}, P_{Z_2}$, respectively. The state noise $S_1$ is known noncausally to the first user and $S_2$ is known noncausally to the second user. The unknown Gaussian noises $Z_1, Z_2$ are independent of the states $S_1, S_2$. The inputs $X_1$ and $X_2$ are subject to power constraints $P_1$ and $P_2$, respectively, i.e. $\mathbb{E}[X_i^2] \leq P_i, i=1,2$.

### III. MAIN RESULTS

In this section, we first derive an achievable rate region for the discrete memoryless TWC with CSI, and then by extending the result to the Gaussian case, we deal with the two-way writing on dirty paper problem.

*Theorem 1:* The capacity region of discrete memoryless TWC with CSI depicted in Fig.1, where $S_i$ is known noncausally to the $i^{th}$ user, $i=1,2$, is lower bounded by the following rate region:

$$\bigcup_{\substack{P_{U_1|S_1} P_{U_2|S_2} \\ X_1 = f_1(U_1, S_1) \\ X_2 = f_2(U_2, S_2)}} \begin{Bmatrix} 0 \leq R_1 < I(U_1; Y_2|U_2, S_2) - I(U_1; S_1|U_2, S_2) \\ 0 \leq R_2 < I(U_2; Y_1|U_1, S_1) - I(U_2; S_2|U_1, S_1) \end{Bmatrix}$$

(2)

*Proof:* We use a random binning argument same as [5]:

Codebook construction:

1) Fix the p.d.f.s $P_{U_1|S_1}$ and $P_{U_2|S_2}$, and two deterministic functions $f_i: \mathcal{U}_i \times \mathcal{S}_i \to \mathcal{X}_i, i=1,2$, where $\mathcal{U}_1, \mathcal{U}_2$ are two arbitrary (finite) sets.

2) Generate randomly $2^{n(R_i+\tilde{R}_i)}$ independent $n$-sequences $u_i^n$ according to the distribution $Pr(u_i^n) = \prod_{t=1}^n P_{U_i}(u_{i,t}), i=1,2$. Label these sequences $u_i^n(w_i, v_i)$, where $w_i \in \{1, ..., 2^{nR_i}\}$ and $v_i \in \{1, ..., 2^{n\tilde{R}_i}\}$.

3) Given $w_i$ and $s_i^n$, let $v_i(w_i, s_i^n)$ be the smallest integer $v_i$ such that $(s_i^n, u_i^n(w_i, v_i)) \in T_\epsilon^n(P_{S_i U_i}), i=1,2$. If such $v_i$ does not exist, set $v_i(w_i, s_i^n) = 1$.

4) The $i^{th}$ encoder, given $w_i$ and $s_i^n$ transmits $x_i^n$, where $x_{i,t} = f_i(u_{i,t}(w_i, v_i(w_i, s_i^n)), s_{i,t}), t=1, ..., n$.

5) The first decoder given $y_1^n$ and having known $(u_1^n(w_1, v_1(w_1, s_1^n)), s_1^n)$ in advance, try to find a pair $(\hat{w}_2, \hat{v}_2)$ such that $(u_1^n(w_1, v_1(w_1, s_1^n)), u_2^n(\hat{w}_2, \hat{v}_2), s_1^n, y_1^n) \in T_\epsilon^n(P_{S_1 U_1 U_2 Y_1})$. If there exist one or more such pair, the first decoder selects one and estimates $\hat{W}_2$ as the corresponding $\hat{w}_2$, and if there is no such pair, sets $\hat{W}_2 = 1$ and declares an error. The decoding role at the second decoder is the same as the first decoder, except the index "1" should be replaced by "2" everywhere.

By a standard analysis of error probability, nearly the same as that one given in [10] for the multiple access channel with correlated CSI, one can show that the average error probability of the code tends to zero provided $n \to \infty$, $\epsilon \to 0$, and also:

$$R_1 < I(U_1; Y_2|U_2, S_2) - I(U_1; S_1|U_2, S_2)$$
$$R_2 < I(U_2; Y_1|U_1, S_1) - I(U_2; S_2|U_1, S_1)$$

(3)

This completes the proof. ∎

*Remark:* In fact, as being induced from the proof of Theorem 1, the rate region given in (2) is achievable for the restricted TWC with CSI. However, next we show that is also optimal for the Gaussian channel in (1). Note that the same result was proved in [2] for the Gaussian TWC without additive interference.

*Two-way writing on dirty paper:* Consider the Gaussian TWC with additive interference defined by (1). It should be noted that the achievability of the rate region (2) was proved for the discrete channel, however one can take the liberty to consider it with the associated input power constraints, also as an achievable rate region for the Gaussian model.

Let $X_1$ and $X_2$ be two independent Gaussian distributed r.v.s with zero mean and variances $P_1$ and $P_2$, respectively. Assume also $X_1, X_2$ are independent of the interferences $S_1, S_2$. Define:

$$U_1 := X_1 + \alpha S_1, \text{ where } \alpha = \frac{cP_1}{c^2 P_1 + P_{Z_2}}$$
$$U_2 := X_2 + \beta S_2, \text{ where } \beta = \frac{bP_2}{b^2 P_2 + P_{Z_1}}$$

(4)

Note that by (4) the Gaussian r.v.s $X_1 - \alpha(cX_1 + Z_2)$ and $cX_1 + Z_2$ are uncorrelated and so independent. Similarly, $X_2 - \beta(bX_2 + Z_1)$ and $bX_2 + Z_1$ are independent. Using this, one can easily check that the following equalities hold:

$$H(U_1|Y_2, U_2, S_2) = H(U_1|Y_2, U_2, S_1, S_2)$$
$$H(U_2|Y_1, U_1, S_1) = H(U_2|Y_1, U_1, S_1, S_2)$$

(5)

So, we can write:

$$I(U_1; Y_2|U_2, S_2) - I(U_1; S_1|U_2, S_2)$$
$$= I(U_1; Y_2, S_1|U_2, S_2) - I(U_1; S_1|U_2, S_2)$$
$$= I(U_1; Y_2|U_2, S_1, S_2) = \frac{1}{2}\log\left(1 + \frac{c^2 P_1}{P_{Z_2}}\right)$$

(6)

and similarly:

$$I(U_2; Y_1|U_1, S_1) - I(U_2; S_2|U_1, S_1) = \frac{1}{2}\log\left(1 + \frac{b^2 P_2}{P_{Z_1}}\right)$$

(7)

By substituting (6) and (7) in (2), we obtain:

*Theorem 2:* The capacity region of the Gaussian TWC with CSI (1), denoted by $\mathcal{C}$, where the interference $S_i$ is known noncausally to the $i^{th}$ user, $i=1,2$, is the same as the capacity when there is no interference and is given by:

$$\mathcal{C} = \begin{cases} 0 \leq R_1 \leq \frac{1}{2}\log\left(1 + \frac{c^2 P_1}{N_2}\right) \\ 0 \leq R_2 \leq \frac{1}{2}\log\left(1 + \frac{b^2 P_2}{N_1}\right) \end{cases} \quad (8)$$

*Proof:* As we obtained, the achievability is induced by the statement of Theorem 1. Further, since the rate region given in (8) is the capacity when there is no interference [2, Sec. VIII], there is no need to prove the converse part. In fact, the rate region given in (8), is also the capacity when the *perfect* CSI, i.e. $(S_1, S_2)$ is available at both users.

*Remark:* As we see from (8), the capacity region of the Gaussian TWC with CSI, same as the channel without interference [2, Sec. VIII], does not depend on the parameters $a$ and $d$, and the feedback from outputs to inputs is completely ineffective. This can be justified as follows: since the $i^{th}$ user knows $X_i(W_i, S_i^n)$ in advance, it will be able to subtract it from its received signal, so the channel can be viewed as the following:

$$Y_1 - aX_1 = bX_2 + S_2 + Z_1$$
$$Y_2 - dX_2 = cX_1 + S_1 + Z_2 \quad (9)$$

In other words, each user perceives the channel as a point-to-point dirty paper channel, and so can transmit by its capacity rate [5].